\documentclass{aa}  

\usepackage{graphicx}
\usepackage{txfonts}

\usepackage{natbib}
\bibpunct{(}{)}{;}{a}{}{,} 

\begin{document}

\title{\object{GCIRS~7}, a pulsating M1 supergiant at the Galactic Centre%
  \thanks{This work relies on interferometric, spectroscopic and
    imaging data obtained at the VLT and VLTI in Cerro Paranal Chile
    between 2003 and 2013. The observations were carried out under the
    program-ids 075.B-0547, 076.B-0259, 077.B-0503, 179.B-0261,
    381.D-0529, 183.B-0100, 087.B-0117, 087.B-0280, 088.B-0308,
    288.B-5040, and 091.D-0682.}\fnmsep%
  \thanks{This work was supported by the French ANR POLCA project
    (Processing of pOLychromatic interferometriC data for
    Astrophysics, ANR-10-BLAN-0511).}}

\subtitle{Physical properties and age}

   \author{T.~Paumard\inst{1}
          \and
          O.~Pfuhl\inst{2}
          \and
          F.~Martins\inst{3}
          \and
          P.~Kervella\inst{1}
          \and
          T.~Ott\inst{2}
          \and
          J.-U.~Pott\inst{4}
          \and
          J.B.~Le~Bouquin\inst{5}
          \and
          J.~Breitfelder\inst{1, 6}
          \and
          S.~Gillessen\inst{2}
          \and
          G.~Perrin\inst{1}
          \and
          L.~Burtscher\inst{2}
          \and
          X.~Haubois\inst{1}
          \and
          W.~Brandner\inst{4}
          }

          \institute{LESIA, Observatoire de Paris, CNRS, UPMC, Université Paris-Diderot,  PSL research university, 5 place Jules Janssen, 92195 Meudon, France\\
              \email{thibaut.paumard@obspm.fr}
              \and
              Max-Planck-Institut f\"ur extraterrestrische Physik, D-85748 Garching, Germany\\
              \email{pfuhl@mpe.mpg.de}
              \and
              LUPM, Universit\'e Montpelier 2, CNRS, Place Eug\`ene Bataillon, F-34095, Montpellier, France
              \and
              Max-Planck-Institut für Astronomie, Königstuhl 17, 69117 Heidelberg, Germany 
              \and
              Univ. Grenoble Alpes, IPAG, F-38000 Grenoble, France   \\
              CNRS, IPAG, F-38000 Grenoble, France
              \and
              European Southern Observatory, Alonso de Córdova 3107, Casilla 19001, Santiago 19, Chile 
             }

   \date{TBD}

 
  \abstract
  {The stellar population in the central parsec of the Galaxy is
    dominated in mass and number by an old (several Gyr) population,
    but young ($6\pm2$~Myr), massive stars dominate the luminosity
    function. The most luminous of these stars is a M1 supergiant,
    {GCIRS~7}.}
  {We have studied {GCIRS~7} in order to constrain the age of the
    recent star formation event in the Galactic Centre and to
    characterise it as a visibility and phase reference for
    observations of the Galactic Centre with the interferometric
    instrument GRAVITY, which will equip the Very Large Telescope
    Interferometer (VLTI) in the near future.}
  {We present the first H-band interferometric observations of
    {GCIRS~7}, obtained using the PIONIER visitor instrument on the VLTI
    using the four 8.2-m unit telescopes. In addition, we present
    unpublished K-band VLTI/AMBER data and build JHKL light-curves
    based on archival data spanning almost 40 years, and measured the
    star's effective temperature using SINFONI integral field
    spectroscopy.}
  {{GCIRS~7} is marginally resolved at H-band, with a
    uniform-disk diameter $\theta_\text{UD}(2013)=1.076\pm0.093$~mas
    ($R_\text{UD}(2013)=960\pm92\;R_\sun$ at $8.33\pm0.35$~kpc). We detect a
    significant circumstellar contribution at K-band. The star and its
    environment are variable in brightness and in size. The
    photospheric H-band variations are well modelled with two periods:
    $P_0\simeq470\pm10$ days (amplitude $\simeq 0.64$~mag) and long
    secondary period $P_\text{LSP}\simeq2700$—$2850$ days (amplitude
    $\simeq 1.1$~mag). As measured from \element[][12]{CO} equivalent
    width, $\left<T_\text{eff}\right>=3\,600\pm195$~K.}
  {The size, periods, luminosity
    ($\left<M_\text{bol}\right>=-8.44\pm0.22$) and effective
    temperature are consistent with an M1 supergiant with an initial
    mass of $22.5\pm2.5\;M_\sun$ and an age of $6.5$--$10$~Myr
    (depending on rotation). This age is in remarkable agreement with
    most estimates for the recent star formation event in the
    central parsec. Caution should be taken when using this star as a
    phase reference or visibility calibrator as it is variable in
    size, is surrounded by a variable circumstellar environment and
    large convection cells may form on its photosphere.}


   \maketitle
%

\section{Introduction}

The central parsec of the Milky Way galaxy is host of a dense star
cluster which is made of a relaxed population of late-type stars
intermixed with a much younger population of luminous, evolved,
massive stars \citep[and references therein]{GenzelEtal2010}. About
one hundred of those young stars seem to be born in one single event
of star formation, presumably within a massive self-gravitating
accretion disk which is hypothesised to have existed at that time
around the super-massive black-hole candidate {Sgr~A*}
\citep[and references therein]{PaumardEtal2006, BartkoEtal2009,
  LuEtal2013}.

{GCIRS~7} is by far the brightest star in the Galactic Centre
(GC) at H and K band. This M1 supergiant \citep{BlumEtal1996nov} has
been first observed in 1966 by \citet{BecklinNeugebauer1968,
  BecklinNeugebauer1975}. Interstellar medium (ISM) features North of
the source are interpreted as the outer layers of the star's
atmosphere being blown away by the central cluster wind in a cometary
tail \citep{SerabynEtal1991, YusefZadehMorris1991}. This star is in
itself an interesting target: it is the brightest of the very few
current-day red supergiant (RSG) stars presumably formed together with
the disk of young stars in the GC \citep{KrabbeEtal1995}. It also
interacts with a wind from either the hot stars or perhaps the
black-hole vicinity itself.

In addition, {GCIRS~7} is an important star for future observations in
the GC. It has often been used as wave-front
reference for infrared adaptive optics systems such as NAOS, and will
be used again for that purpose for interferometric observations with
the 4-telescope beam combiner GRAVITY \citep{EisenhauerEtal2008}. It
will also very likely be used as a fringe-tracker and phase reference
for certain GC observations involving the 1.8-m auxiliary
telescopes (ATs). It is therefore timely to study its interferometric
structure and behaviour.

\cite{PottEtal2008} have observed the star using AMBER at K-band and
MIDI at N-band. They find that the star is marginally resolved at
K-band, more so than its photospheric luminosity would suggest, and
strongly resolved at N-band. They conclude that dust surrounding the
star dominates the visibility in the mid-infrared and has a
non-negligible contribution in the near-infrared.

\citet{BlumEtal1996oct} have shown that the luminosity of {GCIRS~7} has
increased by approximately 0.8, 0.5 and 0.3 magnitudes at J, H and
K-band respectively between 1978 and 1993. \citet{OttEtal1999} have
measured a K-band luminosity decrease of 0.7 magnitudes between 1992
and 1998. Those authors have not been able to determine any particular
regularity in the light-curve and have classified {GCIRS~7} as a
long-period variable (LPV) supergiant.

In Sect.~\ref{sect:obs}, we describe our original and archival data
sets. In Sect.~\ref{sect:results}, we derive stellar parameters such
as the size and effective temperature, interferometric and photometric
variations, age and mass of the star (Table~\ref{tab:params}). We
discuss our findings in Sect.~\ref{sect:discussion}.


\section{Observations}
\label{sect:obs}

\begin{figure}
\includegraphics[scale=0.55]{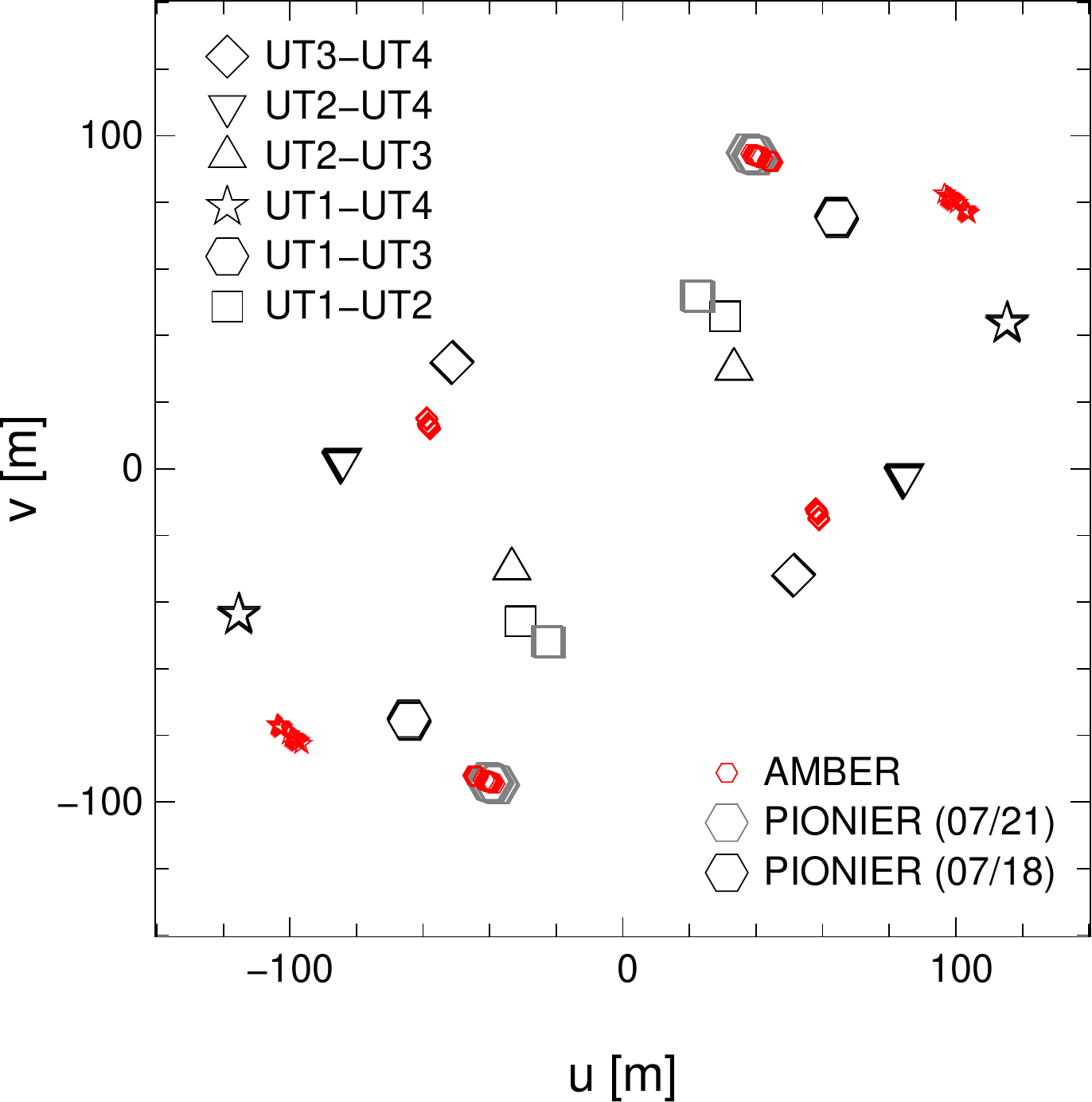}
\caption{Projected baselines ($(u, v)$ coverage) of the PIONIER
  (\emph{large symbols}) and AMBER (\emph{small symbols}) data.}
\label{fig:uvplanes}
\end{figure}

\begin{figure}
\includegraphics[scale=0.333, viewport=-70 0 589 705, clip=true]{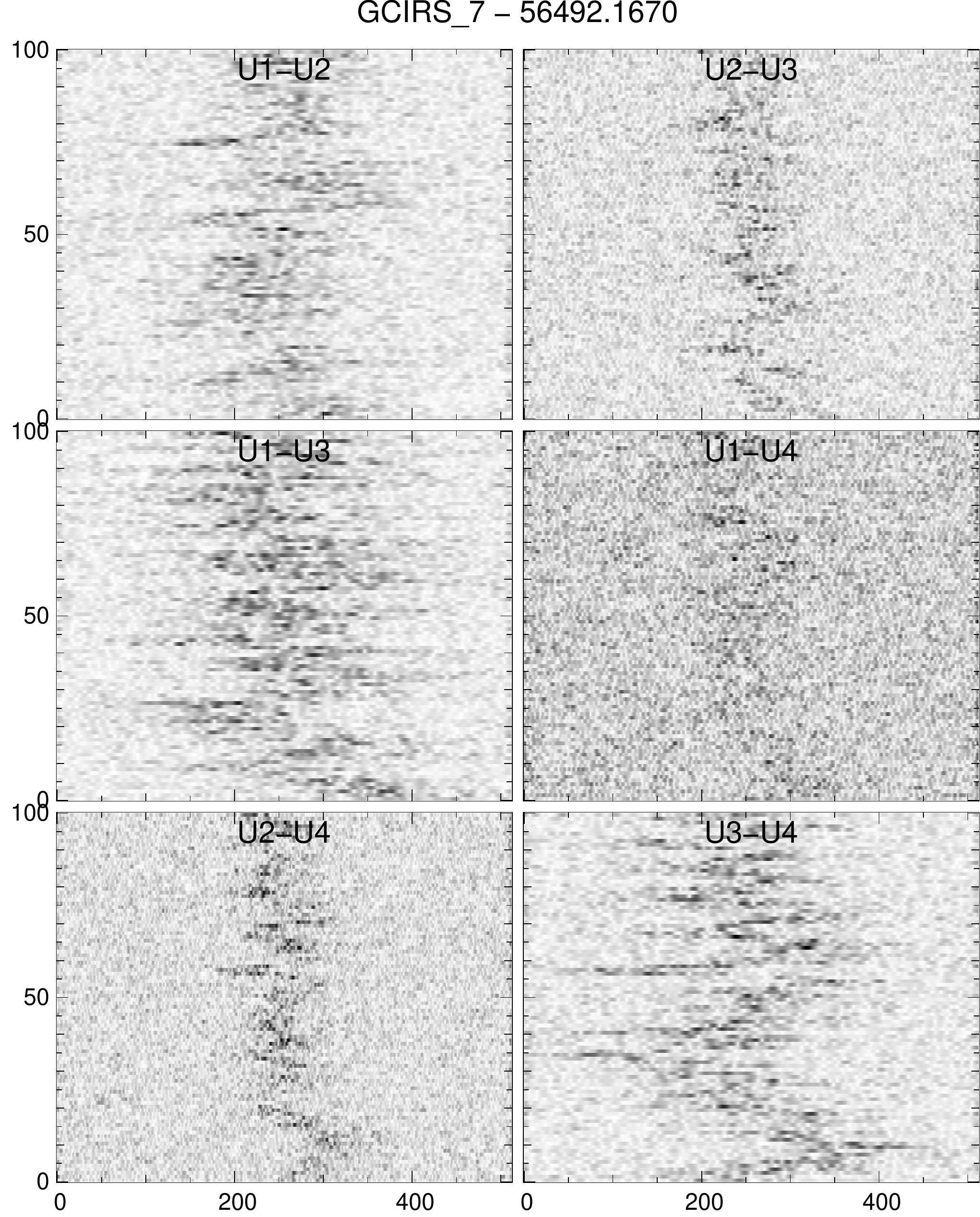}
\caption{Example fringe waterfalls for one exposure. Each panel
  corresponds to one telescope pair. In each panel, the vertical axis
  corresponds to scan number (one scan lasts $\simeq1$~s) and the
  horizontal axis to index within scan (which corresponds to optical
  path difference). In this specific example, fringes are present on
  all baselines.}
\label{fig:waterfall}
\end{figure}

\subsection{Interferometry}
\subsubsection{PIONIER 2013}
\begin{figure}
\includegraphics[scale=0.55]{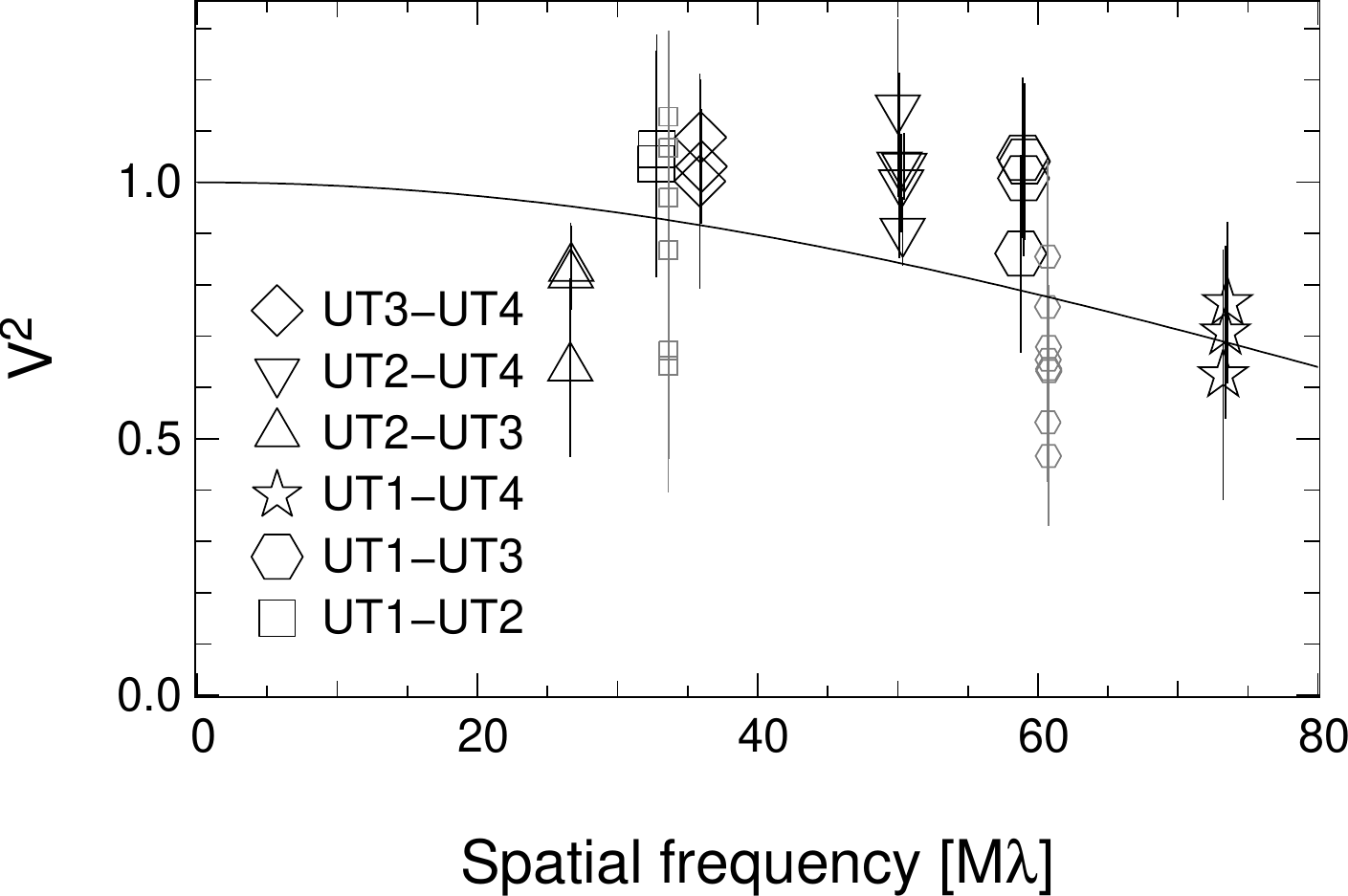}
\caption{2013 H-band PIONIER data, $V^2$ vs. spatial frequency
  in $10^6$/rad or M$\lambda$.
  \emph{Solid line}: uniform-disk model of diameter $1.076$~mas.
  \emph{Larger, black symbols}: first night,
  \emph{smaller, grey symbols}: second night.
}
\label{fig:v2}
\end{figure}

\begin{figure}
\includegraphics[scale=0.55]{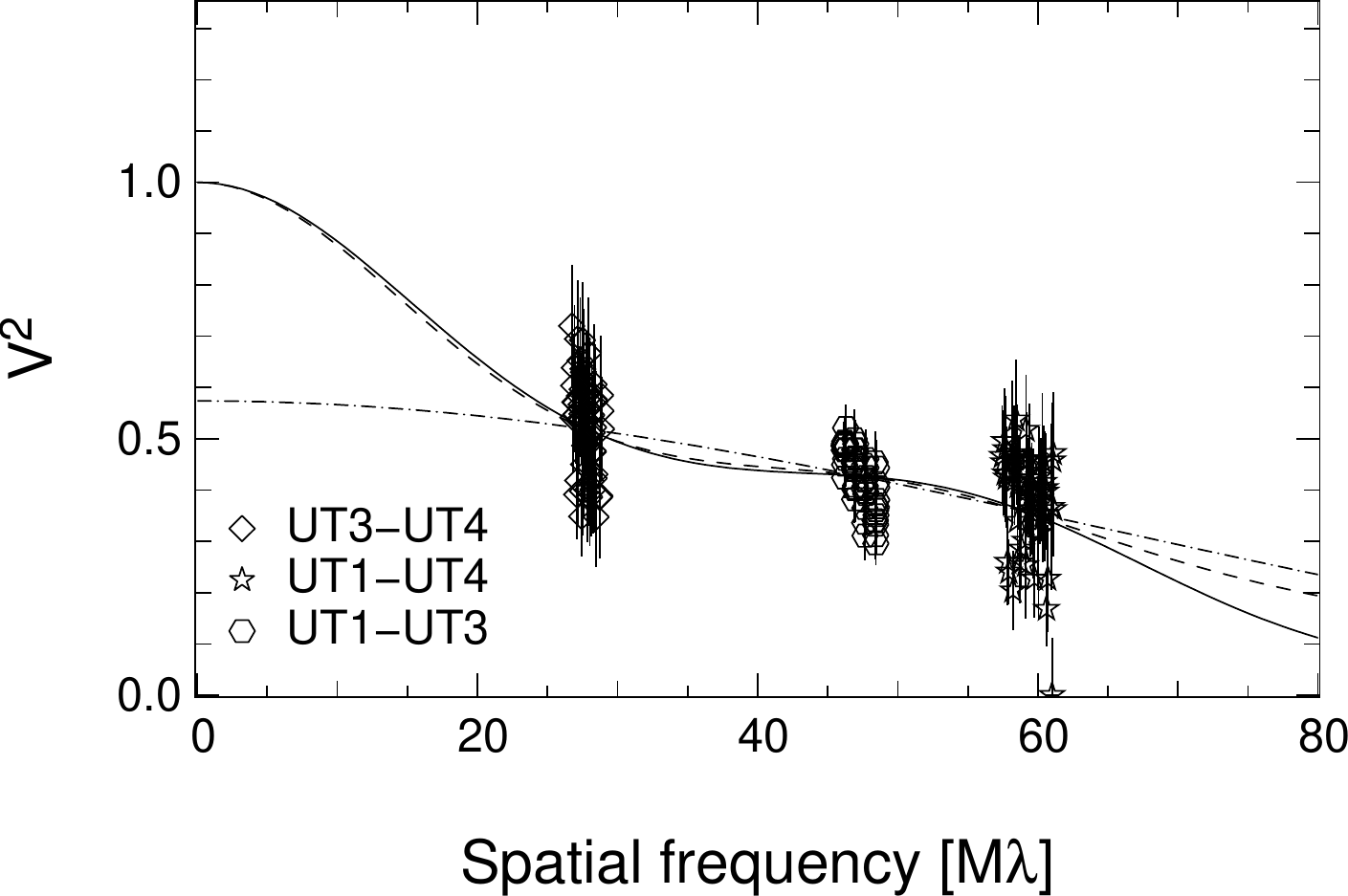}
\caption{2008 K-band AMBER data, $V^2$ vs. spatial frequency in
  $10^6$/rad or M$\lambda$.  \emph{Lines}: several two-components
  models where the star is represented by a uniform disk and the
  environment by one of: \emph{solid line}: a circle, figuring a thin
  shell; \emph{dashed line}: a second uniform disk, figuring a thick
  shell; \emph{dash-dotted line}: a uniform background.}
\label{fig:v2k}
\end{figure}

We have observed the star using PIONIER \citep{LeBouquinEtal2011}
using the four 8.2-m Unit Telescopes (UTs) of the VLTI on July 18th
and 21st, 2013. We used an optical wave-front reference star about
15'' north--East of the target ({USNO-A2.0 0600-28577051}). The $(u,
v)$ coverage is shown on Fig.~\ref{fig:uvplanes}. Since we wanted to
observe fairly faint targets with off-axis adaptive optics, we used
the H-band undispersed mode of the instrument.

The atmospheric conditions on the first night were partly cloudy with
fast coherence time ($\simeq3$~ms).  We were nevertheless able to
record fringes on all 6 baselines (Fig.~\ref{fig:waterfall}) and to
observe a good set of calibration stars.  The second night was
clearer, but coherence time was even shorter ($\simeq1$~ms) so that it
proved impossible to stabilise the fringes, that we were able to
witness by eye, on the detector. We therefore searched for fringes on
a brighter calibrator and then trusted the VLTI delay line model to
apply the right offsets when pointing {GCIRS~7}. We then
recorded scans blindly on the source. Indeed, several scans proved
usable at data reduction time and confirm the visibilities measured on
the first night.

The data have been reduced using the version 2.55 of the PIONIER
pipeline \texttt{pndrs} \citep{LeBouquinEtal2011}. Since our target is
at rather low signal to noise and in order to not introduce biases, we
have decided to select scans where fringe detection is certain from a
visual inspection of the scan waterfalls, but to not filter more
within those scans as additional filtering may lead to biases.

In total, we have been able to obtain data for all six baselines on
the first night and only for U1--U2 and U1--U3 (not at the same time)
on the second night (Fig.~\ref{fig:v2}). Closure phases were also
measured on all three triplets on the first night, all compatible with
$0\degr$ to within the uncertainties, as expected for an
unresolved or marginally resolved source.

\subsubsection{AMBER 2008}

{GCIRS~7} and the calibrator were observed with AMBER in
low-resolution mode and frame exposure time (DIT) of 50~ms, on the
night of May 17, 2008 with the VLTI-UT sub-array UT134. Fringes were
recorded on all three baselines in service mode operation. The $(u,
v)$ coverage is shown on Fig.~\ref{fig:uvplanes}. Seeing was typically
better than 1\arcsec. Due to the high visual extinction towards the
GC, off-axis adaptive optics guiding with MACAO was achieved with the
nearby {USNO-A2.0 0600-28577051}. The visibility calibrator
{HD~164866} was observed close to the target ($\text{dist} =
4.6\degr$). {HD~164866} is with $m_\text{K}=6.5$~mag also of
comparable K-band brightness to {GCIRS~7}.

The data reduction was performed following standard procedures
described in \citet{TatulliEtal2007} and \citet{ChelliEtal2009}, using
the \texttt{amdlib} package, release 3.0.8, and the \texttt{yorick}
interface provided by the Jean-Marie Mariotti Center
(JMMC)\footnote{The calibrated data in the OI-FITS format
  \citep{PaulsEtal2005} will be included in the JMMC database
  \texttt{http://www.jmmc.fr}.}. First, raw spectral visibilities and
closure phases were extracted for all the frames of each observing
file. Then a selection of frames with piston smaller than $15\;\mu$m
was made to achieve higher signal-to-noise ratio (SNR) on the
visibilities and phases. Note that {GCIRS~7} was at the very limit of
sensitivity of AMBER in 2008. Therefore we confirmed to get stable,
calibrated results with two SNR-based frame selections ($20\%$,
Fig.~\ref{fig:v2k}, and $80\%$). The two frame selections within the
piston-limited subset yield the same results, within the stated
uncertainties, which is reassuring. The transfer function was obtained
by averaging the calibrator measurements, after correcting for their
intrinsic diameters. Here again, closure phases are statistically
compatible with $0\degr$.

\begin{figure}
\includegraphics[scale=0.45]{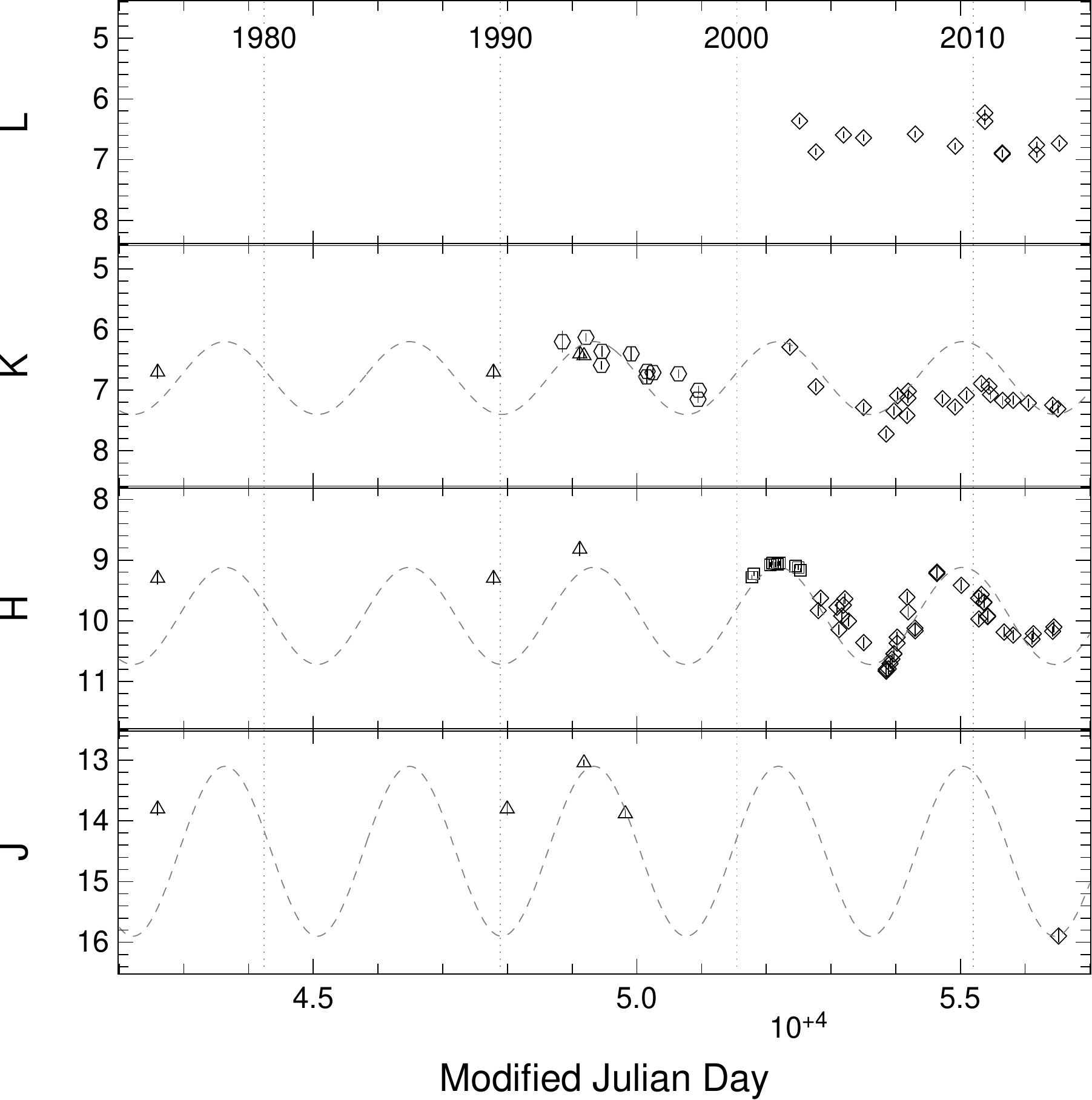}
\caption{J, H, Ks and L-band light-curves.  The oldest data is from
  1975, the latest from June 2013. Statistical uncertainties are
  represented, although they are often smaller than the
  symbol. \emph{Diamonds:} NACO data; \emph{squares:} from
  \citet{PeeplesEtal2007}; \emph{circles:} from \citet[these points
  have been brought up by 0.3 mag]{OttEtal1999}; \emph{triangles:}
  from \citet{BlumEtal1996oct} which includes data from
  \citet{BecklinNeugebauer1975} and
  \citet{DepoySharp1991}. \emph{Dashed curves}: sinusoidal models with
  a period of 2\,850~d.}
\label{fig:phot}
\end{figure}

\subsection{Photometry}

\subsubsection{NACO Imaging Photometry}
\label{sect:obs-naco}

The photometric data were obtained with the adaptive optics camera
NACO \citep{RoussetEtal2003,LenzenEtal2003}. The images were taken
between 2002 and 2013. Most of the images were taken with the H and
Ks-band filters with the 13~$\rm{mas/pixel}$ or the
27~$\rm{mas/pixel}$ camera. Additional L-band and one J-band images
were obtained with the 27\,$\rm{mas/pixel}$ camera. Especially the H
and Ks-band images were selected depending on the saturation level of
{GCIRS~7} in the image. Images with very good seeing and long
integration times had to be removed because they were strongly
affected by saturation. Each selected image was sky-subtracted as well
as bad-pixel and flat-field corrected. In total we used 20 Ks-band, 39
H-band, 21 L-band and one J-band image with temporal spacing between
one day and up to 11 years to construct the light-curve of {GCIRS~7}.

The photometry on the individual images was done by 2D Gaussian fits
to the stars.  As photometric references, the bright early-type stars
{GCIRS~16NE} and {GCIRS~16C} were used. Both stars
showed little or no variability over the recorded time base.  We used
the magnitudes stated by \cite{BlumEtal1996oct} as reference
magnitudes ({GCIRS~16NE}: $m_\text{J}=14.01\pm0.08$,
$m_\text{H}=10.94\pm0.08$, $m_\text{Ks}=9.01\pm0.05$ and
$m_\text{L}=7.56\pm0.09$ / {GCIRS~16C}:
$m_\text{J}=15.20\pm0.08$, $m_\text{H}=11.96\pm0.08$,
$m_\text{Ks}=9.83\pm0.05$ and $m_\text{L}=8.48\pm0.10$).

\subsubsection{Archival Photometry}
\label{sect:obs-archive}

The star {GCIRS~7} is the brightest individual source in the
vicinity of {Sgr~A*} in the near infrared. As such the star was
targeted in many photometric surveys during the last decades. In an
attempt to get a light-curve as long and complete as possible, we used
all published J, H, Ks and L-band data that we were aware of. This
includes data from \citet{BecklinNeugebauer1975},
\citet{DepoySharp1991}, \citet{OttEtal1999}, \citet{BlumEtal1996oct}
and \citet{PeeplesEtal2007}.  By comparing various stars (most notably
GCIRS~16NE and 16C) in the published data sets \citep[in
particular][]{BlumEtal1996oct}, it turned out that the K-band data
from \citet{OttEtal1999} seemed to be offset (brighter) on average by
about 0.3\,mag. This could be related to the choice of the magnitude
reference star. To account for this difference, we added 0.3
magnitudes to the Ott et~al. magnitudes.  The combined light-curve is
shown in Fig.~\ref{fig:phot}.

\subsection{SINFONI Spectroscopy}

\begin{figure}
  \includegraphics[scale=0.5, clip=true, viewport=38 12 490 337]{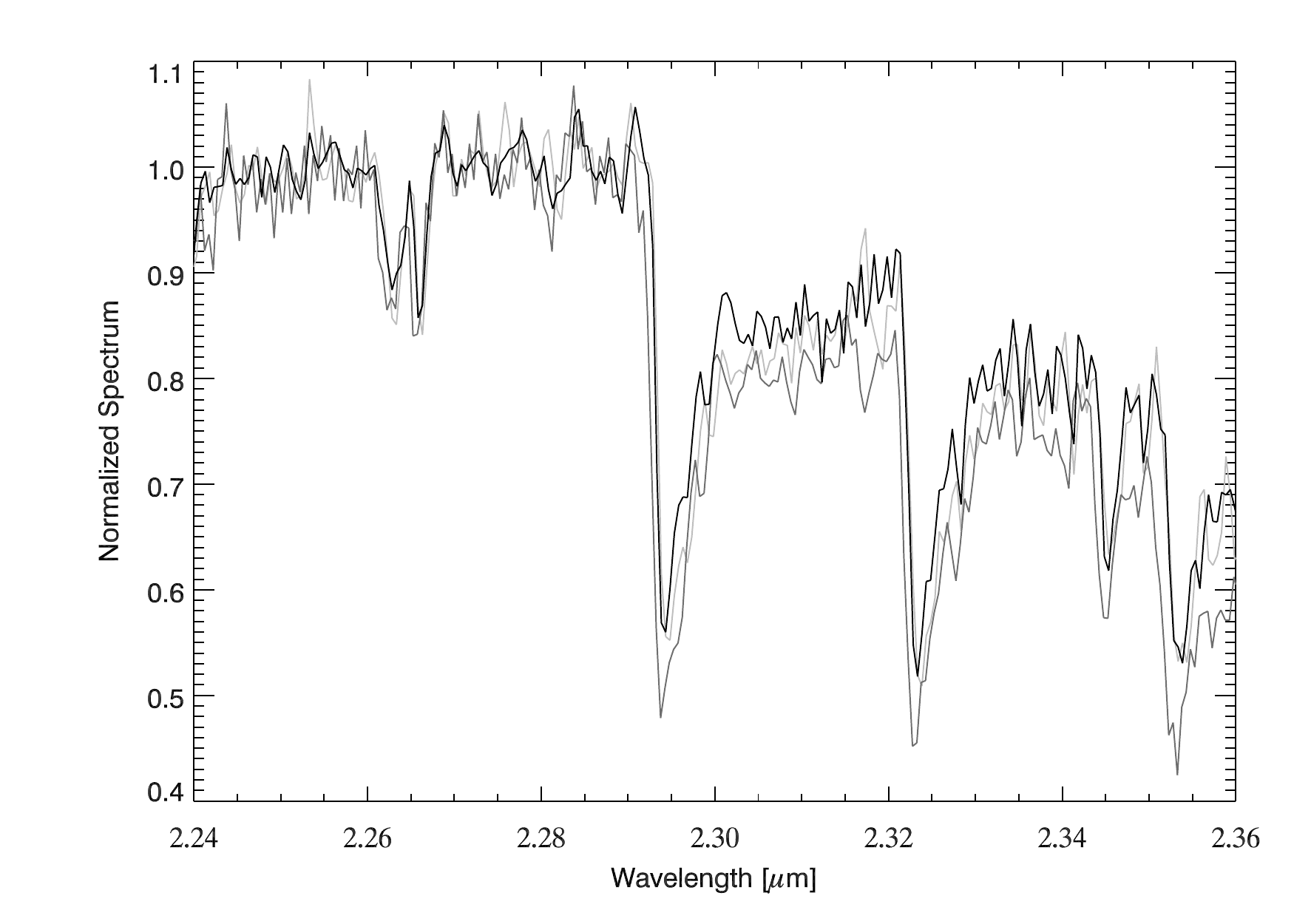}
  \caption{SINFONI spectra. Only the three spectra with a
    resolution of $\simeq2000$ are shown. \emph{Dark gray:}
    $\text{MJD}=52\,738$, \emph{black:} $\text{MJD}=55\,383$; \emph{light
      gray:} $\text{MJD}=56\,558$. }
  \label{fig:spectra}
\end{figure}

\begin{figure}
\includegraphics[scale=0.55]{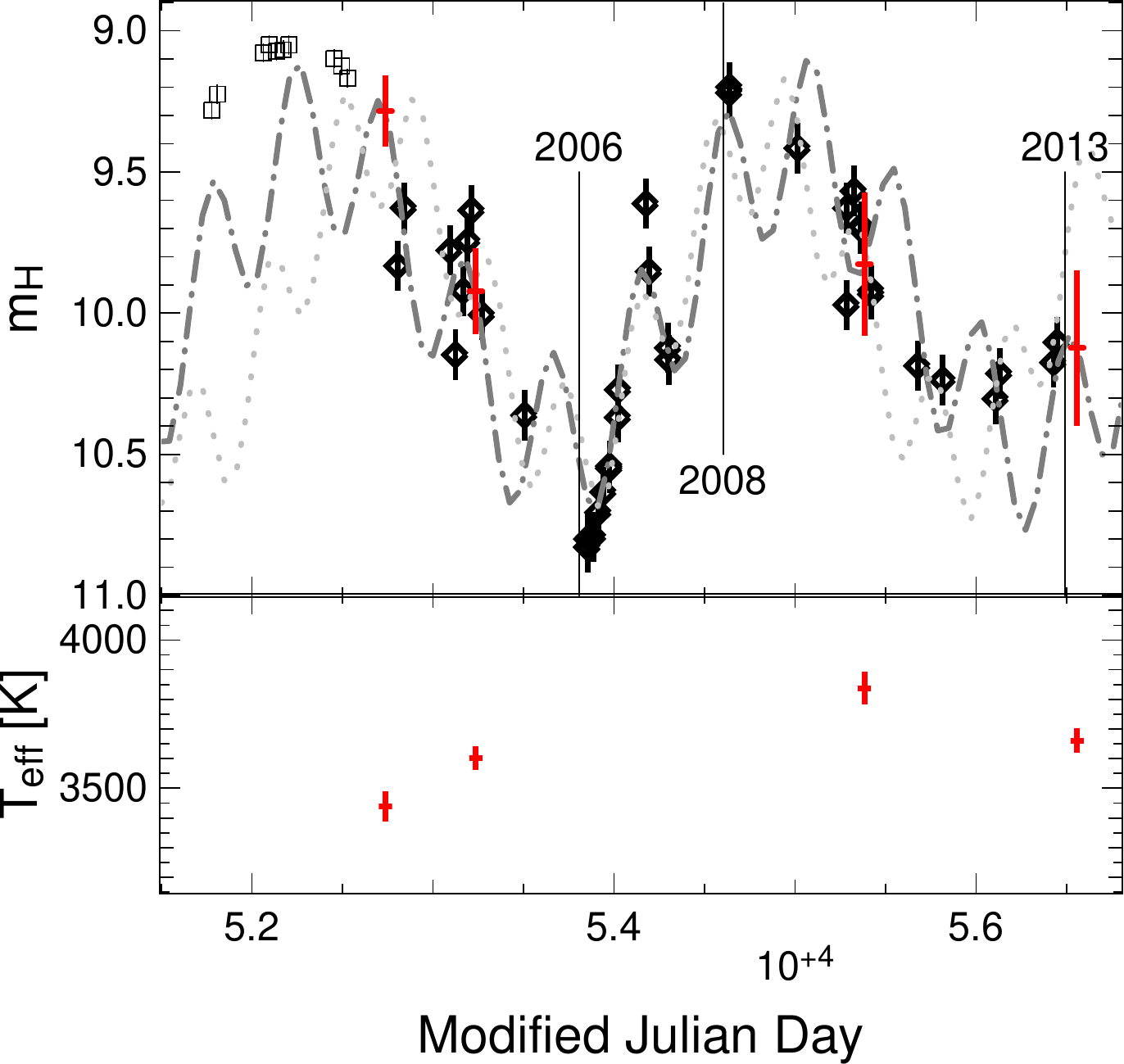}
\caption{\emph{Top:} H-band light-curve after year
  2000. {\emph{Bottom:} effective temperature measurements from
    SINFONI spectra.} \emph{Diamonds:} NACO data (fitted);
  \emph{squares:} \citet[for
  comparison]{PeeplesEtal2007}. \emph{Dash-dotted, dark grey curve}:
  best-fit two-period model. \emph{Dotted, light grey curve}: second
  best-fit model. \emph{Vertical lines}: dates of interferometric
  observations: March 2006 \citep[K-band AMBER]{PottEtal2008}, May
  2008 (K-band AMBER) and July 2013 (H-band
  PIONIER). {\emph{Crosses:} $T_\text{eff}$ from SINFONI spectra
    and the corresponding magnitude estimated using the best fit
    model; the vertical line of each cross indicates the statistical
    uncertainty only (see text, Sect.~\ref{sect:results:teff}).}}
\label{fig:2sine}
\end{figure}

\begin{table}
\caption{Physical parameters}
\label{tab:params}
\centering
\begin{tabular}{lrlc}
  \hline\hline
  Parameter & Value & Uncertainty & Reference \\
  \hline
  \multicolumn{4}{c}{\emph{Direct measurements}}\\
  $m_\text{K}(2013)$         & 7.31    & $\pm0.07$     & Fig.~\ref{fig:phot} \\
  $m_\text{H}(2013)$         & 10.10   & $\pm0.08$     & Fig.~\ref{fig:phot} \\
  $\left<m_\text{K}\right>$  & 6.8     & $\pm0.1$      & Fig.~\ref{fig:phot} \\
  $\left<m_\text{H}\right>$  & 9.93    & $\pm0.03$     & Table~\ref{tab:2sine} \\
  $P_\text{LT}$              & 2\,850  & d             & Fig.~\ref{fig:phot} \\
  $\theta_\text{UD}(2013)$   & $1.076$ & $\pm0.093$ mas & Sect.~\ref{sect:results:size} \\
  $P_0$                     & $470$   & $\pm10$ d     & Sect.~\ref{sect:results:periods} \\
  $P_\text{LSP}$             & 2\,620  & $\pm140$ d    & Sect.~\ref{sect:results:periods} \\
  $\left<\text{EW(CO)}\right>$& 28    & $\pm0.5$      & Fig.~\ref{fig:temperature} \\
  \hline
  \multicolumn{4}{c}{\emph{Indirect measurements}}\\
  $A_\text{K}$               & $3.48$  & $\pm0.09$     & 1 \\
  $A_\text{H}/A_\text{K}$     & 1.73    & $\pm0.03$     & 2 \\
  $R_0$                     & $8.33$  & $\pm0.35$ kpc & 3 \\
  $\rm BC_K$                & $2.84$  & $\pm0.15$    & 4 \\
  $R_\text{UD}(2013)$                 & 960     & $\pm{92}\;R_\sun$  & Sect.~\ref{sect:results:size} \\
  $\left<T_\text{eff}\right>$& 3\,600  & $\pm195$ K    & Table~\ref{tab:temp} \\
  $\left<M_\text{K}\right>$  & $-11.3$ & $\pm0.16$     & Sect.~\ref{sect:absmag}  \\
  $\left<M_\text{H}\right>$  & $-10.7$& $\pm0.2$     & Sect.~\ref{sect:absmag}  \\
  $\left<M_\text{bol}\right>$& $-8.44$ & $\pm0.22$     & Sect.~\ref{sect:absmag}  \\
  $M_\text{K}(2013)$         & $-10.77$ & $\pm0.15$    & Sect.~\ref{sect:absmag}  \\
  $M_\text{H}(2013)$         & $-10.52$& $\pm0.22$     & Sect.~\ref{sect:absmag}  \\
  $M_\text{bol}(2013)$       & $-7.93$ & $\pm0.21$     & Sect.~\ref{sect:absmag}  \\
  $M$                       & 22.5    &$\pm2.5\;M_\sun$& Sect.~\ref{sect:age} \\
  Age                       & 6.5--10 &      Myr      & Sect.~\ref{sect:age} \\
  \hline
\end{tabular}
\tablefoot{
  $m_\text{K}(2013)$, $m_\text{H}(2013)$: K and H-band magnitudes measured in
  June and August 2013 respectively;
  $\left<m_\text{K}\right>$, $\left<m_\text{H}\right>$: baseline magnitudes;
  $P_\text{LT}$: long-term period;
  $\theta_\text{UD}(2013)$: best-fit uniform-disk angular diameter from the
  July 2013 PIONIER data;
  $P_0$, $P_\text{LSP}$: fundamental and long secondary periods found from the
  H-band NACO data;
  $\left<\text{EW(CO)}\right>$: average \element[][12]{CO} equivalent width from the
  SINFONI data;
  $A_\text{K}$, $A_\text{H}/A_\text{K}$: K-band extinction correction and ratio of
  correction at H to K-band;
  $R_0$: distance from the Sun to the GC;
  $\rm BC_K$: bolometric correction for $T_\text{eff}=3\,600$~K;
  $\left<T_\text{eff}\right>$: average effective temperature;
  $R_\text{UD}=(2013)$: uniform-disk physical radius from the 2013 PIONIER data,
  taking limb-darkening into account would yield a $4\%$ larger value;
  $M_\text{bol}(2013)$, $\left<M_\text{bol}\right>$: 2013 and average bolometric
  magnitude;
  $M$: mass.
}
\tablebib{
  (1) \citet{BlumEtal2003};
  (2) \citet{NishiyamaEtal2009};
  (3) \citet{GillessenEtal2009};
  (4) \citet{LevesqueEtal2005}.
}
\end{table}

\label{sect:results}
\begin{figure}
\includegraphics[scale=0.55]{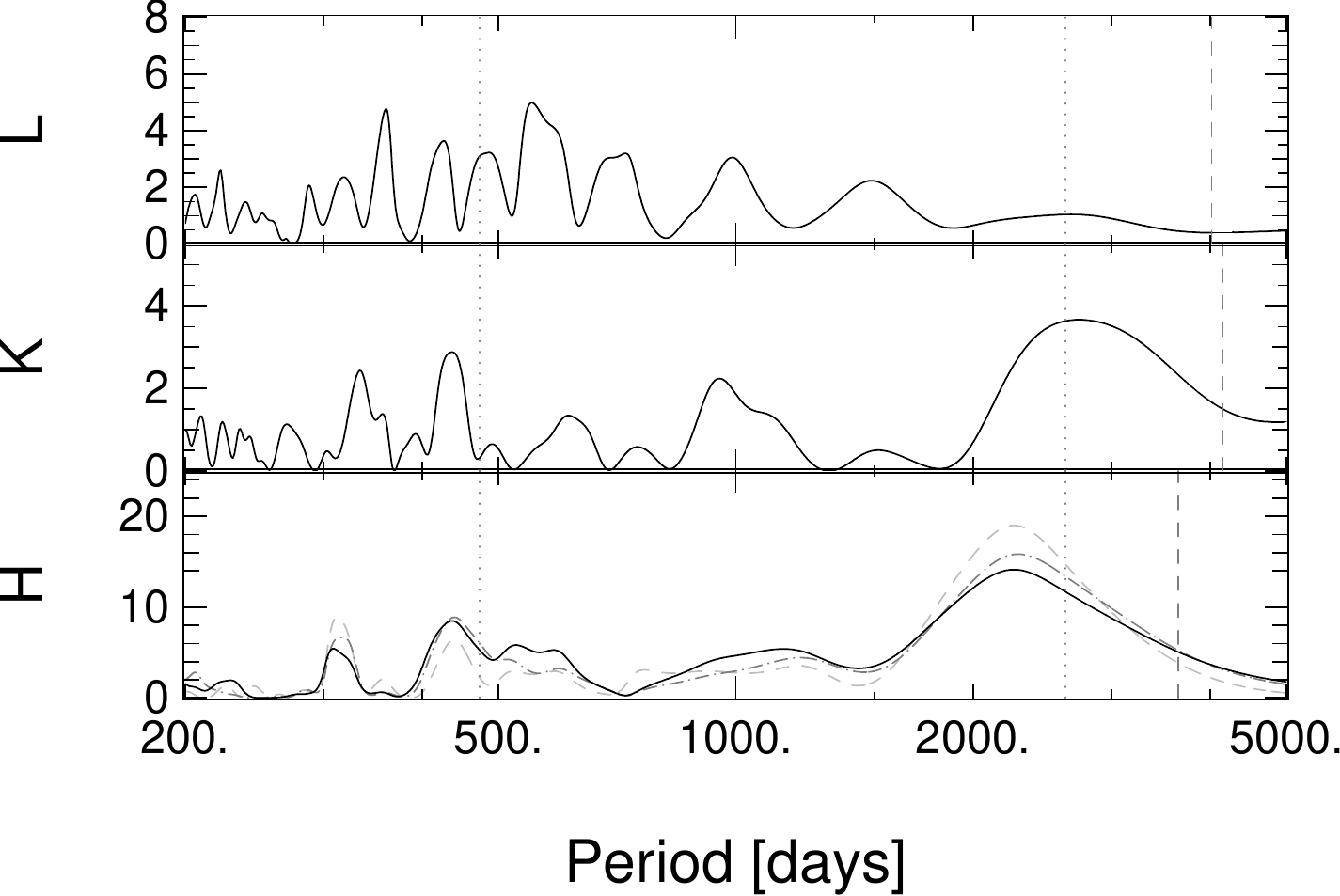}
\caption{Generalised Lomb--Scargle periodograms of the NACO
  light-curves (bottom to top: H, K and L-band). \emph{Black curves}:
  data periodograms. \emph{Light grey, dashed curve}: single-period
  model. \emph{Dark grey, dash-dotted curve:} two-period
  model. \emph{Vertical dotted lines}: best fit
  periods. \emph{Vertical dashed lines}: NACO time coverage in each
  band. The periodograms are normalised so that the expected power for
  pure Gaussian noise is 1.}
\label{fig:periodograms}
\end{figure}

Our spectroscopic data {(Fig.~\ref{fig:spectra})} were obtained
with the adaptive optics assisted integral field spectrograph SINFONI
\citep{EisenhauerEtal2003, BonnetEtal2004}. In total we used {4}
observations obtained in April 2003 {($\text{MJD}=52\,738$)},
August 2004 { ($\text{MJD}=53\,237$), July 2010
  ($\text{MJD}=55\,383$)} and September 2013
{($\text{MJD}=56\,558$)} with pixel scales between $50 \times
100$ and $125 \times 250$\,mas. The data output of SINFONI consists of
cubes with two spatial axes and one spectral axis. Depending on the
plate scale, an individual cube covers $3.2\arcsec \times3.2\arcsec$
or $8\arcsec \times8\arcsec$; the spectral resolving power was about
2\,000 {except for the 2004 run ($R\simeq4\,000$)}. We used the data
reduction SPRED \citep{SchreiberEtal2004,AbuterEtal2006}, including
bad-pixel correction, flat-fielding and sky subtraction. The
wavelength scale was calibrated with emission line gas lamps and
fine-tuned on the atmospheric OH lines. Finally we removed the
atmospheric absorption features by dividing the spectra through a
telluric spectrum obtained in each respective night.

\section{Results}

The various physical parameters derived below, as well as a few
complementary parameters from the literature, are summarised in
Table~\ref{tab:params}.

\subsection {Size}
\label{sect:results:size}

The H-band 2013 visibilities remain very high for spatial frequencies
below $60$~M$\lambda$ (i.e. $B_\perp/\lambda < 6\times10^7$ where
$B_\perp$ is the projected baseline length), except on the shortest
baseline (U2--U3), for which the data have the lowest S/N. A decrease
above $60$~M$\lambda$ ($B_\perp>100$~m at $\lambda=1.68\;\mu$m) is
observed on the first night on the longest baseline (U2--U3), which
also has a fairly low S/N, and on the second night on U1--U3. At any
rate, $V^2 \ga0.7$ even at $73$~M$\lambda$. This points towards a
barely resolved source, if at all.

Indeed, the best-fit uniform-disk diameter is only
$\theta_\text{UD}=1.076\pm0.093$~mas, corresponding to
$R_\text{UD}\simeq960\pm{92}\;R_\sun$ at {$8.33\pm0.35$}~kpc. To estimate a
limb-darkened diameter, we have used the law of \citet{Claret2000} and
the coefficients for $T_\text{eff}=3\,500$~K, $\log(g)=0$, solar
metallicity. The fitted value ($4\%$ larger than $\theta_\text{UD}$,
as expected) is the same within the statistical uncertainties:
$\theta_\text{LD}=1.116\pm0.098\;\text{mas}$. In both cases, the
reduced $\chi^2$ is about 1.9 (the stated error bars are rescaled to
account for the imperfect fits). The reduced $\chi^2$ for
$\theta_\text{UD}=0$ (resp. 2)~mas is $\simeq4$ (resp. 9). Again, this
shows that the source is only moderately resolved.

On the other hand, the K-band 2008 visibilities are of order
$V^2\simeq0.5$ for all baselines. This is a strong indication that up
to $30\%$ of the flux comes from a resolved environment, while
$\ga70\%$ comes from a compact source. We have fitted a few models
where the star is represented by a uniform disk and the environment is
represented by a ring, a second uniform disk, or a uniform
background. In all cases, the photospheric size is in the range
$\theta_\text{UD}1.5$--$2$~mas ($R_\text{UD}=1340$--$1790\;R_\sun$),
the diameter of the circumstellar component is $>5$~mas and the star
accounts for $75$--$85\%$ of the flux. These values should be
considered with caution though, as the $(u, v)$ coverage is not
sufficient given the uncertainties to characterise both the stellar
photosphere and the environment in details.

\subsection{Light-Curve Periodicities}
\label{sect:results:periods}

Figure~\ref{fig:phot} shows the photometric data described in
Sect.~\ref{sect:obs-naco} and \ref{sect:obs-archive}. {GCIRS~7} is
clearly variable at J to L-band, as noted before in the
literature. Our PIONIER H-band observations (modified Julian date
$\text{MJD}\simeq56\,490$) as well as the 2006 AMBER observations
presented in \citet[March 2006:
$\text{MJD}\simeq53\,800$]{PottEtal2008} occurred during a minimum of
the star's brightness. On the contrary, the AMBER K-band observations
we present were acquired near a photometric maximum
($\text{MJD}=54\,603$). The H-band magnitude measured on one NACO
frame in June 2013 (resp. June 2008) was $m_\text{H}(2013)\simeq10.09$
(resp. $m_\text{H}(2008)\simeq9.21$).

For the first time, we are able to show that GCIRS~7 is semi-periodic at
least at H-band with a long-term period of $\simeq$2850 days
(7.8~yr). The H-band light-curve samples very well the last two
pseudo-periods, while \citet{OttEtal1999} K-band data cover the
preceding period. The \citet{BecklinNeugebauer1975} data points are
consistent with the extrapolated sine curves, while the
\citet{DepoySharp1991} points (observed in 1989 and 1990) are too
bright at all wavelengths.

The single periodicity sine-curve model is a fair representation of
the light-curve variations, but departures from this model are
significant. In particular, shorter term variations are seen in the
NACO data (Fig.~\ref{fig:2sine}). Furthermore, RSGs are known to often
exhibit semi-periodic variations. Many stars show a short period
($P_0$, on the order of few 100 days) which is well explained by the
fundamental mode or first overtone of radial pulsation. Other stars
show in addition or instead a longer period ($P_\text{LSP}>1000$~d)
referred to as the long secondary period (LSP), which is less well
explained \citep[see][and references therein]{KissEtal2006}.

Our time sampling is good only for the NACO era, and there it extends
only for a little more than one period. We have constructed the
generalised Lomb--Scargle periodograms
\citep[Fig.~\ref{fig:periodograms},][]{Lomb1976, Scargle1982,
  ZechmeisterKuerster2009} for the H, K and L-band NACO data which
contain data for respectively 37, 20, and 13 distinct dates over 3641,
4142 and 4015 days and attempted simple model fitting of one or two
sine curves.

\begin{table}
  \caption{Best-fit parameters for the two-period model}
  \centering
  \begin{tabular}{lrl}
    \hline\hline
    Parameter & Value & Uncertainty \\
    \hline
    $P_0$ & 470 & $\pm10$ d \\
    $a_0$ & 0.32 & $\pm0.05$ mag \\
    $t_0$ & 52360 & $\pm40$ d \\
    $P_\text{LSP}$ & 2620 & $\pm140$ d \\
    $a_\text{LSP}$ & 0.53 & $\pm0.05$ mag \\
    $t_\text{LSP}$ & 53030 & $\pm70$ d \\
    $\left<m_\text{H}\right>$ & 9.93 & $\pm0.03$ mag \\
    \hline
  \end{tabular}
  \label{tab:2sine}
  \tablefoot{With $t$ the modified Julian date:
    $$m_\text{H}(t)=\left<m_\text{H}\right>+
    a_0 \sin\left(2\pi\frac{t-t_0}{P_0}\right)+
    a_\text{LSP} \sin\left(2\pi\frac{t-t_\text{LSP}}{P_\text{LSP}}\right)$$
  }
\end{table}

\begin{figure}
\includegraphics[scale=0.55]{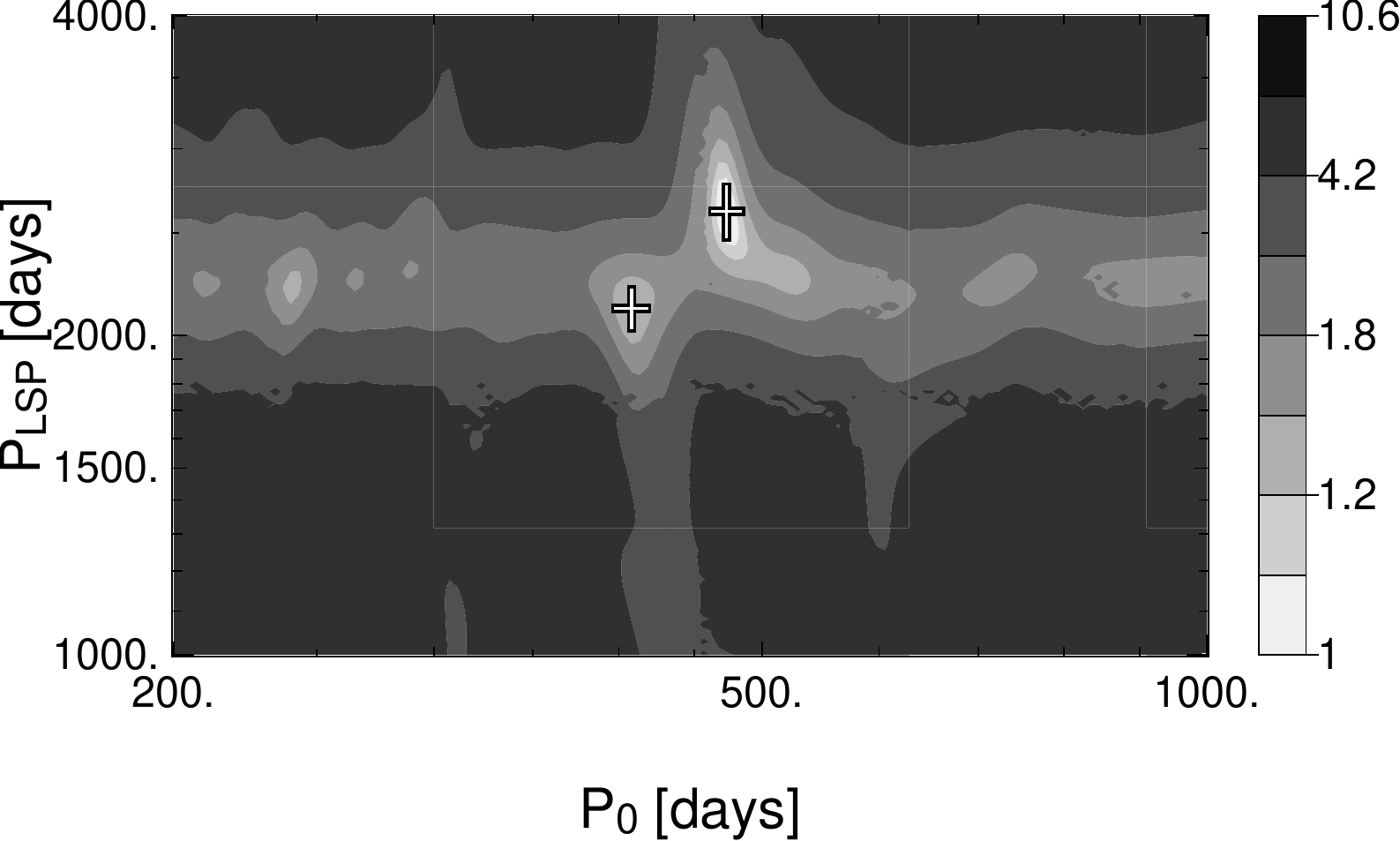}
\caption{$\chi^2$ map for the two-period H-band model, normalised to
  its minimum of 4.68. The best-fit period pairs are marked by
  uncertainty crosses.}
\label{fig:chi2map}
\end{figure}

The $\chi^2$ map of the two-period model (Fig.~\ref{fig:chi2map})
shows that there are only two pairs of periods able to reproduce the
NACO data decently, fairly close to each other:
$(P_0,P_\text{LSP})=(410\pm10\;d, 2120\pm88\;d)$ and $(470\pm10,
2620\pm140)$. They cannot be distinguished using a $\chi^2$ argument
alone.  However, the second pair is closer to the longer term period
seen on Fig.~\ref{fig:phot} and fits the \citet{PeeplesEtal2007} data
points better (Fig.~\ref{fig:2sine}). This is therefore the solution
we consider as best. The $7$ best-fit parameters are listed in
Table~\ref{tab:2sine}.  The amplitudes associated with the two periods
are of the same order: respectively $0.6\pm0.1$ and $1.1\pm0.1$~mag
peak-to-peak for the short and long periods. Although still not
perfect (reduced $\chi^2=4.7$), this two-period model reproduces the
data (Fig.~\ref{fig:2sine}) and their power-spectrum
(Fig.~\ref{fig:periodograms}) much better than the single-period model
(reduced $\chi^2=10.4$).

The K-band light-curve from the NACO era appears almost constant at
$m_\mathrm{K}=7$, except the first point around
$m_\mathrm{K}=6.5$. The most prominent feature of the K-band
periodogram is a broad peak, around $2800$~d, compatible with both the
long-term period ($2850$~d) and the short-term H-band $P_\text{LSP}$
($2620$~d). We tried fitting one or two periods on the K-band data as
well as fitting only the amplitudes for two sine curves at the periods
and phases determined from the H-band data. The reduced $\chi^2$ for
the three different fits remains approximately the same. In
conclusion, the K-band data are in favour of
$P_\text{LSP}\simeq2800$~d, compatible with $P_0$ obtained from the
H-band data, but do not bring further constraints.

Finally, the L-band periodogram is dominated by a band of lines around
$500$~d, which could be the signature of $P_0$ seen in the H-band
data. The data sampling does not allow asserting this firmly, though.

\subsection{Effective Temperature}
\label{sect:results:teff}

\begin{figure}
\begin{center}
\includegraphics[width=\columnwidth]{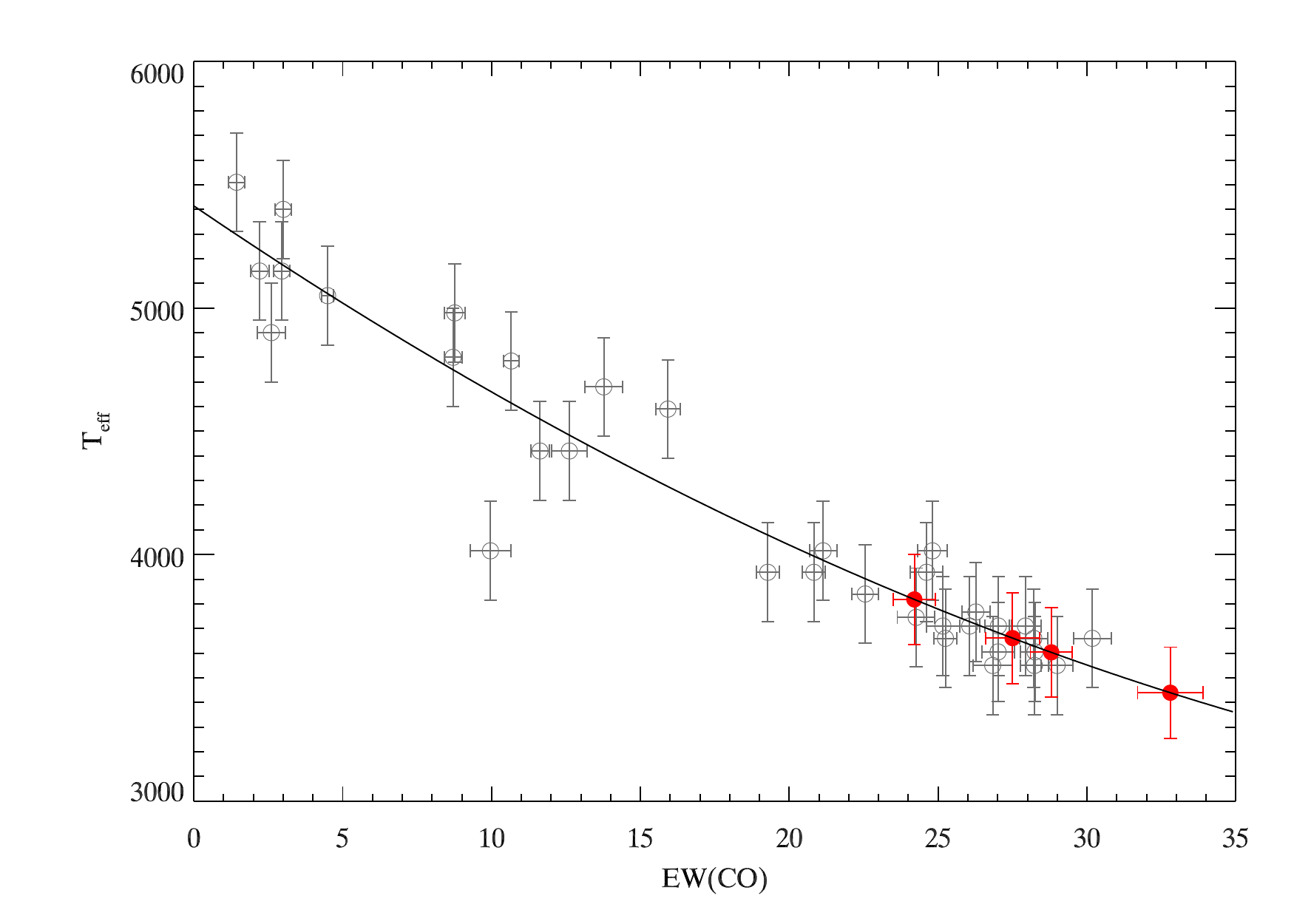}
\caption{CO-$T_\text{eff}$ calibration for RSGs with spectral types
  between G0 and M4. \emph{Open circles}: stars from the IRTF
  library. The temperature uncertainty of the individual stars is
  $200$\,K \citep{AllenCox2000}. \emph{Curve}: best-fit empirical
  relation. \emph{Filled circles}: EW(CO) measurements on {GCIRS~7} and
  derived $T_\text{eff}$.}
\label{fig:temperature}
\end{center}
\end{figure}

Cool stars with temperatures between 3\,000--5\,000\,K show prominent CO
absorption features between 2.29 and 2.40\,$\mu$m.  The CO absorption
strength varies sensitively with temperature, which makes the
so-called CO band heads excellent tracers for the stellar
temperature. Numerous definitions of the CO equivalent width have been
proposed in the literature
\citep[e.g.][]{KleinmannHall1986,FrogelEtal2001}.  Previous studies of
the cool stellar population in the GC such as \cite{BlumEtal2003},
\cite{ManessEtal2007} and \cite{PfuhlEtal2011} determined accurate
\element[][12]{CO}(2,0)--$T_\text{eff}$ calibrations, yet only for
giant stars.  The CO strength however not only depends on the
effective temperature but also on surface gravity. Therefore to get
reliable temperature estimates for the RSG {GCIRS~7}, it is necessary to
determine an adequate \element[][12]{CO}(2,0)--$T_\text{eff}$
calibration for RSGs.

In order to setup such a calibration we used 34 published RSG spectra
($R\simeq$2000) from the IRTF library \citep{RaynerEtal2009}. The
template spectra comprise stars with luminosity classes between Ia to
Ib and spectral types between G0 and M4. We measured the CO strength
according to the definition of \cite{FrogelEtal2001} since their index
proved to be insensitive to biases \citep{PfuhlEtal2011}.  The
effective temperature of the library stars was derived based on their
spectral type. For spectral types between G0 and K0 we used
temperatures from \cite[p. 152]{AllenCox2000}. In case of later
spectral types K1 to M5, we used the revised calibration from
\cite{LevesqueEtal2005}. The result is shown
Fig.~\ref{fig:temperature}. As expected, the CO strength clearly
correlates with the stellar temperature. We used a 2nd-order
polynomial fit get an empirical relation between EW(CO) and effective
temperature.  The best-fit relation is:
\begin{equation}
T_\text{eff} = 0.669 \cdot {\rm CO}^2 - 82.139 \cdot {\rm CO} +5414 \rm ~~[K]
\end{equation}

The intrinsic scatter of the template stars with respect to the
best-fit temperature is 180~K. This is roughly consistent with the
stated temperature uncertainty of the spectral types of
$\sim$200\,K\citep[p. 153]{AllenCox2000}. All available temperature
measurements are summarised in Table\,\ref{tab:temp}. Although the
measurements used various techniques, they all agree very well. {
  In the following, we use the mean of the measurements
  $\left<T_\text{eff}\right> =3\,600\pm195$~K as an estimate of the
  average temperature of the star. The uncertainty is the mean of the
  individual uncertainties and is dominated by systematic errors. The
  {root-mean-square} scatter between the individual measurements
  is only {116}~K.}
\begin{table}
\caption{Effective temperature of {GCIRS~7} \label{tab:temp}}
\centering
\begin{tabular}{ccc}
\hline\hline
{Date (MJD)} &
{$T_{\rm eff}$ [K]\footnotemark{}}&
{Reference}\\
\hline
56\,558 & $3\,661\pm185$ & \dots \\ 
55\,383 & $3\,818\pm183$ & \dots \\ 
53\,237 & $3\,603\pm183$ & \dots \\
52\,738 & $3\,439\pm185$ & \dots \\
52\,426 & $3\,650\pm150$ & 1 \\
51\,332 & $3\,600\pm200$ & 2 \\
50\,648 & $3\,430\pm240$ & 3 \\
49\,122 & $3\,600\pm230$ & 4 \\
{\bf Average} & $\mathbf{3\,600\pm195}$ & \dots \\
\hline
\end{tabular}
\tablefoot{All uncertainties include systematic errors. For
  the SINFONI data set, 180~K have been added in quadrature to the
  statistical uncertainties to this effect.}
\tablebib{
  (1) \cite{CunhaEtal2007};
  (2) \cite{DaviesEtal2009};
  (3) \cite{BlumEtal2003};
  (4) \cite{CarrEtal2000}.}
\end{table}

{Even though the systematic errors are too large to derive a
  consistent temperature curve, EW(CO) does change measurably between
  SINFONI runs (Fig.~\ref{fig:temperature}), and the colours also vary
  quite significantly. We investigate the temperature variations by
  considering only the statistical uncertainties in the four SINFONI
  $T_\text{eff}$ estimates, assuming that the systematic effects
  affect those four points in the same fashion. We have also estimated
  the H-band magnitude of the star at the corresponding dates using
  our best fit model (Table~\ref{tab:2sine}). We used uncertainties on
  the fit parameters to derive uncertainties on these
  magnitudes. Those temperatures, magnitudes and the corresponding
  uncertainties are shown Fig.~\ref{fig:2sine}.

  To assess the significance of the variation in $T_\text{eff}$
  between SINFONI runs, we first take the average of these four
  values: $\left<T_\text{eff}\right>_\text{SINFONI}=3\,629\pm18$~K.
  The departure from this value for each date is respectively
  $-190\pm46$, $-26\pm36$, $209\pm49$ and $32\pm36$~K. This is on
  average a $2.5\sigma$ departure.
}

\subsection{Absolute Magnitude and Bolometric Luminosity}
\label{sect:absmag}
The intrinsic K and H-band luminosity of {GCIRS~7} can be derived from
the average observed $\left<m_\text{K}\right>=6.8\pm0.1$ and
$\left<m_\text{H}\right>=9.93\pm0.03$, the measured K-band extinction
$A_\text{K}=3.48\pm0.09$ \citep{BlumEtal2003} and
$A_\text{H}/A_\text{K}=1.73\pm0.03$, \citep{NishiyamaEtal2009,
  Fritz2013} and the distance modulus $d=14.6\pm0.09$ \citep[assuming
$R_0=8.33\pm0.35$~kpc,][]{GillessenEtal2009}. The absolute, dereddened
K and H-band magnitudes are therefore
$\left<M_\text{K}\right>=-11.3\pm0.16$ and
$\left<M_\text{H}\right>=-10.7\pm0.2$. The bolometric K-band
correction for a star with $T_{\rm eff}=3\,600\pm195\,$K is $\rm
BC_{K}=2.84\pm0.15$ \citep{LevesqueEtal2005}. Thus the (average)
absolute bolometric magnitude of {GCIRS~7} is $\left<M_{\rm
    bol}\right>=-8.44\pm0.22$, corresponding to a luminosity of
$\left<L\right>=1.86\pm0.4 \times 10^5\,L_{\sun}$. The position of
{GCIRS~7} in the Hertzsprung--Russell (HR) diagram can be seen in
Fig.~\ref{fig:tracks}.

Likewise, the 2013 values of the K and H-band and bolometric absolute
magnitudes are $M_\text{K}(2013)=-10.77\pm0.15$,
$M_\text{H}(2013)=-10.52\pm0.22$ and $M_\text{bol}(2013)=-7.93\pm0.22$
respectively.

\begin{table}
  \caption{Variations of $T_\text{eff}$, $M_\text{H}$ and $R_\text{LD}$
    from the SINFONI spectra and NACO H-band photometry\label{tab:Rt}}
  \centering
  \begin{tabular}{cr@{}lr@{}lr@{}l}
    \hline\hline
    MJD & \multicolumn{2}{c}{$T_\text{eff}$ [K]} &   \multicolumn{2}{c}{$M_\text{H}$} & \multicolumn{2}{c}{$R_\text{LD}$ [$R_\sun$]} \\
    \hline
    56\,558 & 3\,661&$\pm32$ & 10.1 &$\pm0.3$ &  950&$\pm120$ \\
    55\,383 & 3\,838&$\pm45$ &  9.8 &$\pm0.2$ & 1030&$\pm120$ \\
    53\,237 & 3\,603&$\pm31$ &  9.9 &$\pm0.1$ & 1068&$\pm71$ \\
    52\,738 & 3\,439&$\pm42$ &  9.3 &$\pm0.1$ & 1524&$\pm85$ \\
    \hline
  \end{tabular}
  \tablefoot{In this table, only statistical uncertainties
    are considered. For instance, when propagating also systematic
    sources of error, the uncertainty on $R_\text{LD}$ reaches
    $200\;R_\sun$ for 2003.}
\end{table}

{To estimate the bolometric magnitude of the star at the time of
  the SINFONI observations, we lack a well-established H-band
  bolometric correction law. However, using $T_\text{eff}$,
  $m_\text{H}$, and the relations in \citet{KervellaEtal2004}, we can
  derive the radius of the star for these four epochs. From these
  radii and the temperatures, we can directly get the bolometric
  magnitude at each SINFONI observation. In this estimation, we are
  only interested in the statistical uncertainties and therefore do
  not consider the uncertainties on $A_\text{K}$,
  $A_\text{H}/A_\text{K}$ and $R_0$. The values are listed in
  Table~\ref{tab:Rt}.}

\subsection{Age and initial mass}
\label{sect:age}
\begin{figure}
\centering
\includegraphics[width=\columnwidth]{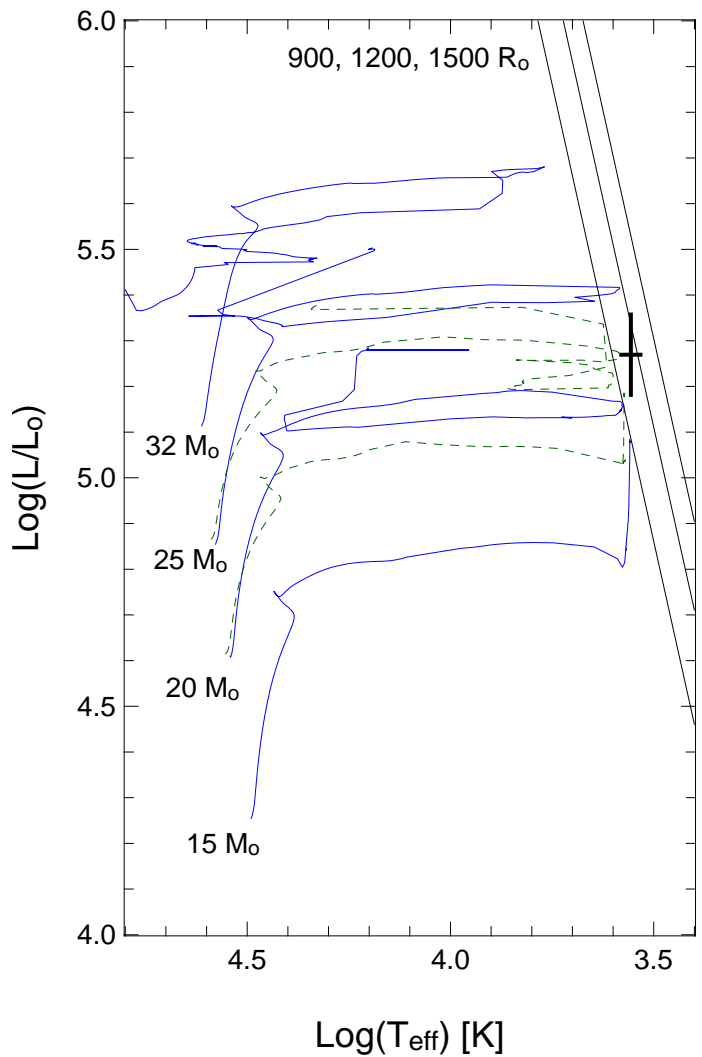}

\caption{Position of \object{GCIRS~7} in the H-R diagram (cross) compared with
  evolutionary tracks of stars with initial masses between $M_{\rm
    ini}=15-32\,M_{\sun}$ with rotation (blue solid) and for $M_{\rm
    ini}=20-25\,M_{\sun}$ without rotation (dashed green) from
  \cite{EkstroemEtal2012}.  The solid black lines denote the
  theoretical luminosities of stars with radii of $R=900, 1100$ and
  $1300 \,R_{\sun}$ as function of effective temperature.}
\label{fig:tracks}
\end{figure}
In order to estimate the initial mass and the age of {GCIRS~7}, we
rely on the recent evolutionary tracks from
\citet[Fig.~\ref{fig:tracks}]{EkstroemEtal2012}. Their tracks include
stellar rotation and assume a solar metallicity. This seems to be a
reasonable choice since {GCIRS~7} is known to have a metallicity close
to solar \citep{CarrEtal2000,CunhaEtal2007}.  {GCIRS~7} falls between
the evolutionary tracks of $M_{\rm ini}=20\,M_{\sun}$ and $M_{\rm
  ini}=25\,M_{\sun}$ stars with and without rotation.  Stars with
initial masses of $M_{\rm ini}=20\,M_{\sun}$ reach the RSG phase after
about 8\,Myr without rotation and 10\,Myr with rotation and last there
for about 0.2\,Myr. Stars with initial masses of $25\,M_{\sun}$ reach
the RSG phase after about 6.5\,Myr without and after 8\,Myr with
rotation and remain there for a few ten thousand years.  Stars with
initial masses of $15\,M_{\sun}$ (RSG age 14-15\,Myr) do not reach the
observed luminosity. Unfortunately evolutionary tracks with a finer
mass sampling are {available only through an interpolation
  tool\footnote{\url{http://obswww.unige.ch/Recherche/evoldb/index/Interpolation/}}.
  GCIRS~7 lies on the far right end of the tracks, and the
  interpolation seems to cut some of this temperature extremum.  The
  authors of the interpolation tools note that it can not be relied
  upon for unstable phases}. However the available tracks {(as
  well as the interpolated tracks)} are roughly equally spaced in
luminosity. Based on that we estimate that {GCIRS~7} originates from a
star with initial mass of $22.5\pm 2.5 \,M_{\sun}$ \citep[in good
agreement with][$22\ M_\sun$]{CunhaEtal2007}. Depending on the initial
rotation of the star, the age of {GCIRS~7} is in the range
6.5--10~Myr.

\section {Discussion}
\label{sect:discussion}

\subsection{On the size of {GCIRS~7} and its environment}
\label{sect:discuss:size}
\label{sect:circstell}

{The effective temperature and radius of 74 Galactic RSGs has been
  determined by \citet{vanBelleEtal2009}. From their equation (1), we
  can estimate $(V-K)_0=4.5\pm0.3$ for GCIRS~7. This part of their
  Fig.~5 ($(V-K)_0>3.5$, $R>500\;R_\sun$) is mostly occupied by stars
  from \citet{LevesqueEtal2005}, with which GCIRS~7 is quite
  consistent.}

Using $T_\text{eff}=3\,600$~K, $m_\text{H}(2013)=10.10$, $A_K=3.48$
and $A_H/A_K=1.73$ (Table~\ref{tab:params}), we compute a
limb-darkened diameter of 1.10~mas using the relations given in
\citet{KervellaEtal2004}. The K-band relation with
$m_\text{K}(2013)=7.3$ yields a very similar value. This is well
within $1\sigma$ of our 2013 measurement.  The angular diameter we
fitted on the 2008 K-band AMBER data ($1.5$--$2$~mas) is affected by
degeneracy with the circumstellar environment. However, it is
consistent with the photometry: the H-band relation in
\citealt{KervellaEtal2004} yields $\theta_\text{LD}=1.68$,
$R_\text{LD}=1500\;R_\sun$ for $m_\text{H}(2008)=9.21$, still assuming
$T_\text{eff}=3\,600$~K. {In addition, this diameter is also
  consistent with the value we determine spectro-photometrically for
  the 2003 epoch, at which the star was approximately as bright as in
  2008 (Table~\ref{tab:Rt}, Fig.~\ref{fig:2sine}).}

The question which arises is whether this apparent change in radius is
real. Indeed, no radius pulsation has been detected so far on similar
RSGs. A change from $\simeq1\,000\;R_\sun$ to $1\,500\;R_\sun$
translates to $\Delta R/\left<R\right>\simeq40\%$, which is somewhat
larger than the maximum amplitude observed in cepheids
\citep{Tsvetkov1988}. In addition, $R\simeq1\,500\;R_\sun$ is the size
of the largest known RSGs \citep{LevesqueEtal2005,
  ArroyoTorres2013}. {Our interferometric measurements hint at a
  change of diameter. This change is confirmed by our analysis of the
  available spectroscopy and photometry. However, the systematic
  uncertainty of $T_\text{eff}$ translates into a large
  ($\simeq200R_\sun$) systematic uncertainty on the radius. In
  addition, the spectro-photometric estimates assume that
  the H-band local extinction was constant throughout the 10
  years. Furthermore, the largest estimate
  ($R_\text{LD}(2003)=1\,500\pm200\;R_\sun$, including systematic
  errors) is based on an extrapolation of the NACO light-curve}.  In
conclusion, the unique best {direct measurement} for the size of
GCIRS~7 is our PIONIER 2013 H-band measurement, which is not subject
to degeneracy with the circumstellar environment, {the star is
  probably truly variable in size,} and $\theta_\text{UD}$ is unlikely
to reach values in excess of 1.7~mas ($R\simeq1\,500\;R_\sun$).

This confirms the interpretation in \citet{PottEtal2008}, where the
loss of visibility was eventually attributed to a circumstellar
contribution. However, the visibility measured in 2006 at K-band in
\citet{PottEtal2008} was actually quite high compared to our 2008
measurement ($V^2\simeq0.8$ at $\simeq20$~M$\lambda$). Since the star
was very faint at that time, one would naively expect the
circumstellar emission to dominate and therefore the visibility to be
quite low. This was not the case. This means that not only the star,
but also its environment has varied between 2006 and 2008. The 2006
measurement can be understood with either (or both) of the following
hypotheses:
\begin{itemize}
\item the circumstellar emission was fainter in 2006 than in 2008,
  so that the ratio of one over the other has not changed much;
\item the circumstellar emission was more compact when the star itself
  was smaller and fainter, such that the visibility of the extended
  component was larger in 2006 than in 2008.
\end{itemize}

A circumstellar contribution could also explain why the K-band (and
L-band) magnitude has not varied much in the last decade: photospheric
variations at longer wavelengths are perhaps diluted in a significant,
steady circumstellar emission. This does not explain, however, why the
star appears to have been fainter at K-band in recent years compared
to what it was at the end of the XXth century. Obscuration by newly
formed circumstellar material is not a very tempting explanation, as
it would have impacted H-band photometry even stronger than K-band
photometry. In addition, this extended component would need to be
emission rather than scattering, since scattering is more efficient at
shorter wavelengths. Black-body emission from dust could have a
significant contribution at K-band while remaining basically
undetectable from our H-band PIONIER data. Finally, $A_\text{K}$ is
larger by $\simeq1$~mag in the direction of {GCIRS~7} compared to the
rest of the central parsec \citep{SchoedelEtal2010}, which points
towards a fairly thick circumstellar environment.

As mentioned in the introduction, ISM features North of
  GCIRS~7 have been interpreted as the outer layers of the star's
  atmosphere being blown away by the central cluster wind in a
  cometary tail \citep{SerabynEtal1991, YusefZadehMorris1991}. Another
  evidence of this cluster wind is provided by the very clear,
  large-scale bow-shock South-East of GCIRS~3
  \citep{ViehmannEtal2005}. The fact that our size measurement is
  quite consistent with the bolometric magnitude and effective
  temperature of the star tends to show that the interaction with the
  cluster wind does not affect the photospheric appearance of the star
  nor, presumably, its mass-loss rate. Further investigations,
  however, should be able to measure the actual distribution
  of circumstellar material around the star.

\subsection{On the variability of {GCIRS~7}}
\label{sect:discussion:variability}
\begin{figure}
  \includegraphics[scale=0.55]{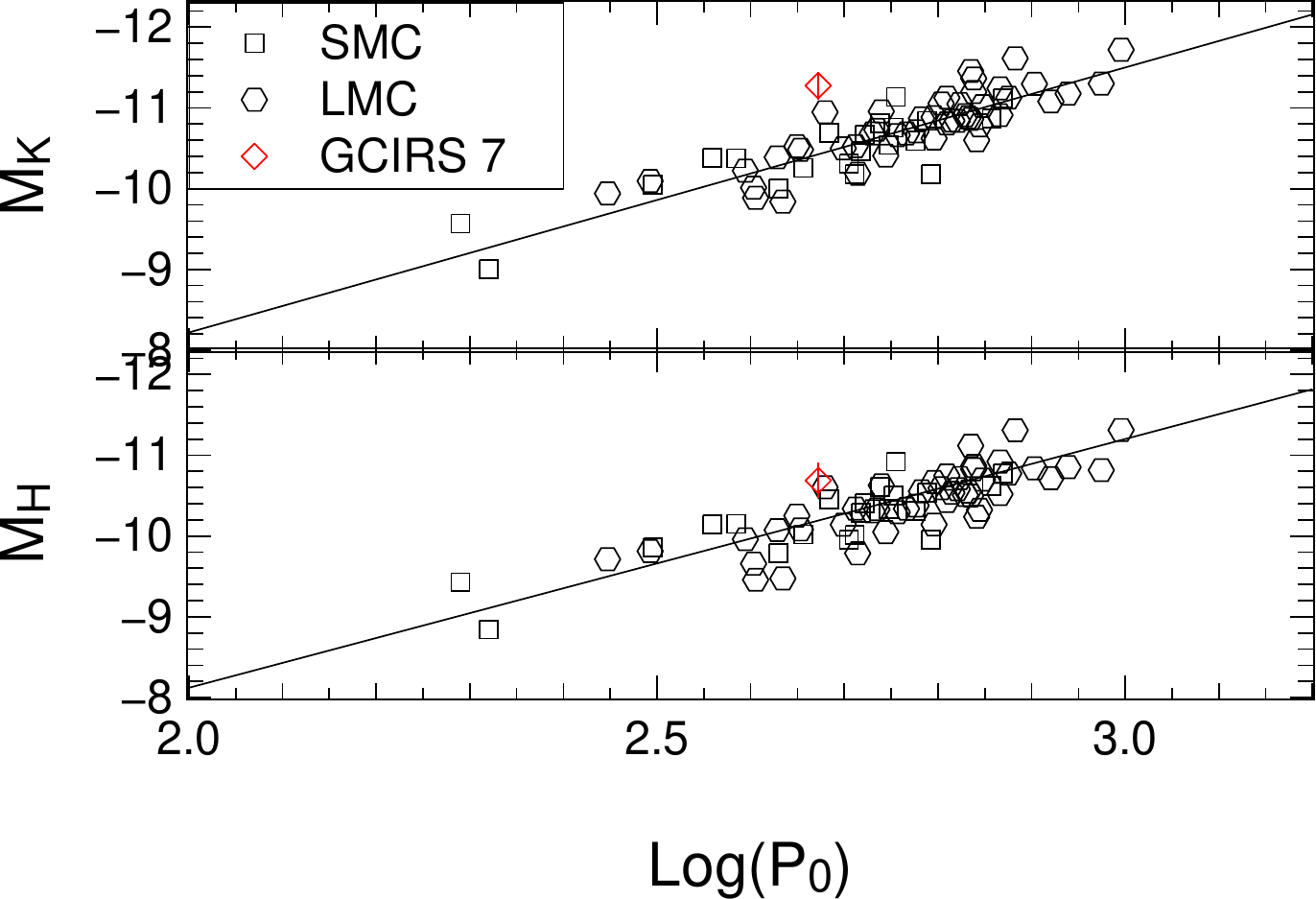}
  \caption{Period-Luminosity relation for RSGs. \emph{Diamond with
      vertical error bar}: {GCIRS~7}. \emph{Squares and circles}: SMC
    and LMC stars used in \citet{YangJiang2012}. \emph{Line}: best-fit
    law for the LMC$+$SMC \citep[Table~5]{YangJiang2012}.}
  \label{fig:PL}
\end{figure}

The short, primary period of RSGs follows a fairly tight
period-luminosity (P-L) relation \citep[][and references
therein]{YangJiang2012}. Figure~\ref{fig:PL} presents {GCIRS~7}
($M_K=-11.3\pm0.16$, $M_H=-10.7\pm0.2$, $P_0=470\pm10$~d, see
Table~\ref{tab:params}) on the P-L relationship for RSGs in the SMC
and LMC. Given the uncertainty on the absolute magnitudes of {GCIRS~7}
and the intrinsic scatter around the best-fit relation, the agreement
with the P-L relation is quite good, especially at H-band. The K-band
magnitude of the star appears to be $\simeq0.1$~mag too bright, which
could be due to the circumstellar contribution discussed
Sect.~\ref{sect:circstell}.

In contrast to the short, primary period of RSGs, there is as of now
no P-L relation established for the LSP. However, \citet{KissEtal2006}
have noticed a relation between the LSP, mass and radius of RSGs. They
considered $W=P(M/M_\sun)(R/R_\sun)^{-2}$, which ``is the natural form
of the pulsation constant if the oscillations are confined to the
upper layers of the envelope \citep{GoughEtal1965}''. They computed
the average of $W$ for 14 RSGs with $P > 1000\;\text{d}$ and found
$\left < W \right > =0.082\pm 0.03$. Taking $P=2620$—$2850\;\text{d}$,
$\left<R\right>=1000\ R_\sun$ and $M=22.5\;M_\sun$
(Table~\ref{tab:params}, Sect.~\ref{sect:discuss:size}), we find for
{GCIRS~7} $W=0.059$—$0.064$, within $1\sigma$ of the above mentioned
average.

\subsection{{GCIRS~7} as an interferometric calibrator for GRAVITY}

{GCIRS~7} is the only star in the central parsec bright enough to be
used as fringe-tracker reference with GRAVITY operating on the VLTI
auxiliary telescopes (ATs). Furthermore, this star being bright and
close from the central black-hole candidate {Sgr~A*}, it will be
tempting to use it as a visibility calibrator whenever observing an
object in the central parsec with GRAVITY, be it with the UTs or with
the ATs.

However, GRAVITY will operate in the K-band. The K-band visibilities
of {GCIRS~7} are affected by a significant, variable circumstellar
contribution (Sect.~\ref{sect:circstell}). It will therefore be
difficult to use this star as a visibility calibrator. Nevertheless,
assuming the circumstellar geometry is simple enough (pending further
investigations), one could perhaps use the star as a local calibration
proxy. At any rate, the star must be considered interferometrically
variable on the time scale of about one month, a fraction of the
$\simeq470$~d short period that we measured.

In addition, the size and nature of {GCIRS~7} imply that convection
cells may form in its atmosphere and impact the appearance of the
photosphere \citep[see e.g.][]{HauboisEtal2009,
  FreytagHoefner2008}. \citet{ChiavassaEtal2011} have shown that the
photometric wobbling for a Betelgeuse-like star is of order 0.1~AU,
corresponding to $\simeq10\;\mu$as at the distance of the Galactic
Centre. This is of the same order as the expected astrometric accuracy
of GRAVITY itself. This will limit the astrometric precision of
studies of the nuclear star cluster using GCIRS~7 as astrometric
reference. This limitation may be alleviated by systematically
calibrating the astrometric zero point by observing one or several
smaller or more stable stars.

\subsection{On the age of {GCIRS~7} and the recent star
  burst}
We list various physical parameters in Table~\ref{tab:params}, which
are all consistent with each other. This gives good confidence in the
age we derive for {GCIRS~7}: $6.5$ to $10$~Myr, depending on
rotation.

RSGs are in a rare state. The ratio of red to blue
supergiants is approximately 0.4--$0.5\%$
\citep[e.g.][]{PfuhlEtal2011}, so that there must be approximately 200
blue supergiants in the Galactic Centre with the same age as
GCIRS~7. There is a unique, well known population of $>100$ hot,
massive stars in the central parsec, and the age we derive for GCIRS~7
is consistent the age estimated by several authors for this population
($2$--$7$~Myr, \citealt{GenzelEtal2003}; $6\pm2$~Myr
\citealt{PaumardEtal2006, BartkoEtal2009}). On the other hand,
\citet{LuEtal2013} derived a $95\%$ confidence interval for a cluster
age of 2.5 to 5.8 Myr. This claim is not compatible with the age we
derive for GCIRS~7.

\citet{LuEtal2013} relied essentially on the ratio of Wolf--Rayet (WR)
to OB stars. This method has been recognised to be strongly dependent
on the prediction of evolutionary tracks
\citep{SchaererVacca1998}. The evolution of a star beyond the main
sequence, especially in the WR phase, is very sensitive to numerical
and physical prescriptions such as overshooting and mass loss
rate. For an illustration, see e.g. \citet{MartinsPalacios2013}. In
addition, the definition of a WR star does not rely on the same
criteria in evolutionary models and in spectroscopic surveys. Finally,
it is crucial to have complete sample to correctly evaluate the
observed ratio of WR to OB stars.

This is why in \citet{PaumardEtal2006} we did not rely only on this
indicator. In addition, we looked for the turn-off in the HR diagram
\citep[Fig.~12]{PaumardEtal2006}, i.e. the position in the HR diagram
of the most massive stars in the main sequence. The age of {GCIRS~7}
is a now third independent age indicator. Taken together, these
indicators point towards a cluster age in the range $6$--$8$~Myr. This
age is actually permitted by the error bars of Fig.~3, right panel, of
\citet{LuEtal2013}, especially if one allows for a slightly flatter
initial mass function than their preferred slope.

The simultaneous presence of both RSG and WR stars in a massive
cluster presumably formed in a single burst of star formation is
rare. For that to happen, the cluster must be young enough for the
most massive stars to still be present (in the WR phase). At the same
time, it must be old enough so that stars in the mass range
10--25~M$_{\sun}$ have evolved to the RSG phase. The fact that we are
able to explain the presence of OB, WR stars and the RSG {GCIRS~7}
using a single isochrone is a very strong indication that the central
cluster is in this peculiar age range where all types of stars are
present.

\section{Conclusions}

We have obtained interferometric fringes on {GCIRS~7} with PIONIER on
the six UT baselines of the VLTI at H-band. We have been able to
measure the photospheric size of the star:
$\theta_\text{UD}(2013)\simeq1.076\pm0.093\;\text{mas}$ translating to
$R_\text{UD}(2013)\simeq960\pm{92}\ R_\sun$ at $R_0=8.33\pm0.35\
\text{kpc}$. K-band AMBER observations obtained in 2008 show a
significant ($\simeq20\%$) circumstellar excess, and hint at
photospheric and circumstellar variations. {Photospheric
  temperature and size variations are confirmed by spectroscopy and
  photometry.}

In addition, we have reanalysed near infrared light-curves of the star
and found a prominent, long period of $2620$—$2850$ days and a shorter
period of $470\pm10$ days. The global peak-to peak variation has been
$\simeq2$~mag at H-band and $\simeq1.6$~mag at K-band during the last
40~yr. The size we measure is quite consistent with the luminosity
($\left<M_\text{bol}\right>=-8.44\pm0.22$) and effective temperature
($T_\text{eff}=3\,600\pm195$~K) of the star, and the short and long
periods are in good agreement with the fundamental and long secondary
period respectively for a RSG that size and that mass
($22.5\pm2.5\;M_\sun$, Sect.~\ref{sect:age}). The physical parameters
we derive are listed in Table~\ref{tab:params}.

Future observations of the Galactic Centre nuclear star cluster may be
performed with GRAVITY using {GCIRS~7} as a fringe-tracker
reference. Given the variable nature of this star and its environment,
it may not be a stable phase reference at the expected accuracy of
GRAVITY ($10\;\mu$as) for durations of more than a few
weeks. Therefore, it may be useful to calibrate the astrometric
zero-point on smaller or more stable stars regularly when using
GCIRS~7 as a phase reference. Likewise, its usability as a (secondary)
K-band visibility calibrator depends on whether the environment
morphology is simple enough, pending further investigations.

Finally, the age we derive ($6.5$--$10$~Myr) is a confirmation of the
cluster age ($6\pm2$~Myr) we determined \citep{PaumardEtal2006,
  BartkoEtal2009} and contradicts recent claims for a younger cluster
\citep[$2.5$--$5.8$~Myr,][]{LuEtal2013}.

\begin{acknowledgements}
  We thank Myriam Benisty for help with the AMBER data reduction. This
  research has made use of Jean-Marie Mariotti Center products: the
  \texttt{AMBER data reduction package}\footnote{Available at
    http://www.jmmc.fr/amberdrs} and the
  \texttt{LITpro}\footnote{LITpro software available at
    http://www.jmmc.fr/litpro} service \citep{TallonBoscEtal2008}
  co-developed by CRAL, LAOG and FIZEAU, as well as of NASA's
  Astrophysics Data System.  We acknowledge financial support from the
  ``Programme National de Physique Stellaire" (PNPS) of CNRS/INSU,
  France.
 
\end{acknowledgements}


\bibliographystyle{aa} 
\bibliography{main}{}

\end{document}